\documentclass[11pt]{article}

\usepackage{amsmath}
\usepackage{amsthm}
\usepackage{amsfonts}
\usepackage{amssymb}
\usepackage{citesort}
\usepackage{a4wide}
\usepackage{graphicx}

\def\e{\mathrm{e}}
\def\i{\sqrt{\!-\!1}}
\def\Define{:=}
\def\definE{=:}
\def\nn{\nonumber}
\def\id{\mathrm{id}}
\def\End{\mathrm{End}}

\newtheorem{thm}{Theorem}[section]
\newtheorem{prop}[thm]{Proposition}

\title{\bf \mathversion{bold}
The $L(\mathfrak{sl}_{2})$ symmetry of \\[2mm]
the Bazhanov-Stroganov model 
associated with \\[2mm]
the superintegrable chiral Potts model
\mathversion{normal}}

\author{
Akinori Nishino\footnote{E-mail address: %
nishino@iis.u-tokyo.ac.jp}
and
Tetsuo Deguchi$^{1}$\footnote{E-mail address: %
deguchi@phys.ocha.ac.jp}\\
\,\\
Institute of Industrial Science, The University of Tokyo,\\[2mm]
4--6--1 Komaba, Meguro-ku, Tokyo 153--8505, Japan \\
\,\\
$^{1}$Department of Physics, Ochanomizu University,\\[2mm]
2--1--1 Ohtsuka, Bunkyo-ku, Tokyo, 112--8610, Japan
}

\date{}

\begin{document}

\maketitle
\begin{abstract}
\setlength{\baselineskip}{15pt}
The loop algebra $L(\mathfrak{sl}_{2})$ symmetry 
is found in a sector of the nilpotent Bazhanov-Stroganov model.
The Drinfeld polynomial of a $L(\mathfrak{sl}_{2})$-degenerate 
eigenspace of the model is equivalent to 
the polynomial~\cite{Albertini-McCoy-Perk-Tang,Baxter,McCoy-Roan,%
Albertini-McCoy_2,Dasmahapatra-Kedem-McCoy,Tarasov,Baxter_2} 
which characterizes a subspace with the Ising-like spectrum
of the superintegrable chiral Potts model.
\end{abstract}

\section{Introduction}

The chiral Potts model is a two-dimensional solvable lattice model 
whose Boltzmann weights lie on a curve of genus greater than
one~\cite{YMPTL,Baxter-Perk-AuYang,Perk-AuYang}.
Albertini, McCoy, Perk and Tang numerically found 
a special point on the curve where the spectra of 
the transfer matrix fit to an Ising-like simple 
form~\cite{Albertini-McCoy-Perk-Tang}.
The model at the special point is called the 
``superintegrable'' chiral Potts (SCP) model,
and its free energy and interfacial tension
are explicitly calculated~\cite{Baxter}.

In order to discuss Onsager's approach to the SCP model~\cite{Onsager},  
let us consider the $\mathbb{Z}_{N}$-symmetric Hamiltonian 
introduced by 
von Gehlen and Rittenberg~\cite{vonGehlen-Rittenberg}: 
\begin{align}
 \label{eq:Zn-symmetric}
 &H_{\mathrm{vGR}}=A_{0}+k^{\prime}A_{1}
  =\frac{4}{N}\sum_{i=1}^{L}\sum_{m=1}^{N-1}
  \frac{1}{1-\omega^{-m}}\big(Z_{i}^{2m}
  +k^{\prime} X_{i}^{-m}X_{i+1}^{m}\big).
\end{align}
Here, the operators 
$Z_{i},X_{i}\in\End((\mathbb{C}^{N})^{\otimes L})$ 
are defined by
\begin{align}
\label{eq:ZX-op}
&Z_{i}(v_{\sigma_{1}}\otimes\cdots\otimes 
   v_{\sigma_{i}}\otimes\cdots\otimes v_{\sigma_{L}})
 =q^{\sigma_{i}}v_{\sigma_{1}}\otimes\cdots\otimes 
   v_{\sigma_{i}}\otimes\cdots\otimes v_{\sigma_{L}},\nn\\
&X_{i}(v_{\sigma_{1}}\otimes\cdots\otimes 
   v_{\sigma_{i}}\otimes\cdots\otimes v_{\sigma_{L}})
 =v_{\sigma_{1}}\otimes\cdots\otimes 
   v_{\sigma_{i}+1}\otimes\cdots\otimes v_{\sigma_{L}},
\end{align}
for the standard basis $\{v_{\sigma}|\sigma=0,1,\cdots,N-1\}$
of $\mathbb{C}^{N}$ and under the periodic boundary conditions 
$Z_{L+1}=Z_{1}$ and $X_{L+1}=X_{1}$. 
Here and hereafter we assume odd $N$ for simplicity, 
and set $q=\e^{\frac{2\pi}{N}\i}$ and $\omega=q^{2}$ when $q^N=1$.
The Hamiltonian $H_{\mathrm{vGR}}$ is derived from  
the expansion of the SCP transfer matrix with respect to 
 the spectral parameter.
The $A_{0}$ and $A_{1}$ in~\eqref{eq:Zn-symmetric} satisfy
the Dolan-Grady conditions
$[A_{i},[A_{i},[A_{i},A_{1-i}]]]=%
16[A_{i},A_{1-i}], (i=0,1)$~\cite{Dolan-Grady} 
and generate the Onsager algebra (OA)~\cite{Onsager}. 
The Hilbert space $(\mathbb{C}^{N})^{\otimes L}$ 
is decomposed into the direct sum of  finite-dimensional irreducible 
representations of OA. In each of the subspaces 
the energy spectra of $H_{\mathrm{vGR}}$ 
fit to the Ising-like form \cite{Davies}:
\begin{align}
\label{eq:energy-form_SCP}
  E\!=\!\alpha\!+\!k^{\prime}\beta
  \!+\!2\sum_{i=1}^{n}m_{i}
  \sqrt{1\!+\!2k^{\prime}\cos\theta_{i}\!+\!k^{\prime 2}},\quad
  m_{i}\in\{-l_{i}, -l_{i}\!+\!2,\cdots,l_{i}\}.
\end{align}
Here, $\alpha, \beta\in\mathbb{R}$ and 
$\theta_{i}\in\mathbb{R}/2\pi\mathbb{R}$ are such parameters 
that are determined not only by the OA approach. 

The parameters $\alpha, \beta$ and $\theta_{i}$
in the spectra \eqref{eq:energy-form_SCP} are determined by 
the functional relations of the SCP transfer matrix.
Introduce~\cite{Albertini-McCoy-Perk-Tang,Baxter,McCoy-Roan,%
Albertini-McCoy_2,Dasmahapatra-Kedem-McCoy,Tarasov,Baxter_2}
\begin{align}
\label{eq:CP-poly}
&P_{\mathrm{CP}}(\xi^{N})
 =\omega^{-p_{b}}\sum_{j=0}^{N-1}
  \frac{(1-\xi^{N})^{L}(\xi\omega^{j})^{-p_{a}-p_{b}}}
  {(1-\xi\omega^{j})^{L}
   F_{\mathrm{CP}}(\xi \omega^{j})F_{\mathrm{CP}}(\xi\omega^{j+1})}, 
\end{align}
and call it the chiral Potts polynomial.
Here $F_{\mathrm{CP}}(\xi)\Define\prod_{i=1}^{R}
 (1+\xi u_{i}\omega)$ and  $\{u_{i}\}$ satisfy 
the following coupled nonlinear equations: 
\begin{align}
\label{eq:CP-Bethe-eq}
& \left(\frac{u_{i}+\omega^{-1}}{u_{i}+\omega^{-2}}\right)^{L}
 =\omega^{-p_{a}-p_{b}-1}\prod_{j=1\atop j(\neq i)}^{R}
  \frac{u_{i}-u_{j}\omega^{-1}}{u_{i}-u_{j}\omega},
\end{align}
and $p_{a}$ and $p_{b}$ are chosen so that $P_{\mathrm{CP}}(0)$
is finite and non-zero.
The coupled nonlinear equations \eqref{eq:CP-Bethe-eq}
give the pole-free conditions for $P_{\mathrm{CP}}(\zeta)$.
If the polynomial $P_{\mathrm{CP}}(\zeta)$ is factorized as 
$P_{\mathrm{CP}}(\zeta)
=\prod_{i=1}^{n}(1-\zeta_{i}^{-1}\zeta)^{l_{i}}$ with 
distinct zeros $\zeta_{1},\ldots,\zeta_{n}\in\mathbb{C}$,
the parameters $\{\theta_{i}\}$ in~\eqref{eq:energy-form_SCP} 
are determined through $\zeta_{i}=-\tan(\theta_{i}/2)$ and
the dimensions of the corresponding representation space of OA
are given by $\prod_{i=1}^{n}(l_{i}+1)$~\cite{Davies}.
The  chiral Potts
polynomial $P_{\mathrm{CP}}(\zeta)$ is first obtained 
in the special cases: the sector $R=0$~\cite{Baxter} 
and the case $N=3$~\cite{Albertini-McCoy-Perk-Tang}.
The expression \eqref{eq:CP-poly} for $P_{\mathrm{CP}}(\zeta)$ 
with general $N$ and $R$ is given by Baxter \cite{Baxter_2}.
We remark that the results 
in~\cite{Albertini-McCoy-Perk-Tang,Baxter,Tarasov,Baxter_2} 
imply $l_{i}=1$ for all $i$, that is, 
the dimensionality of the OA representation space is 
$2^{\mathrm{deg} P_{\mathrm{CP}}(\zeta)}$
where $\mathrm{deg} P_{\mathrm{CP}}(\zeta)$ denotes
the degree of $P_{\mathrm{CP}}(\zeta)$.

The chiral Potts polynomial $P_{\mathrm{CP}}(\zeta)$
plays the central role in 
the spectra \eqref{eq:energy-form_SCP}. 
However, its definition is still nontrivial at least algebraically.   
In fact, it is based on the functional relations for the transfer 
matrices of chiral Potts model~\cite{Baxter-Bazhanov-Perk}, 
which are rather complicated,  
and any mathematical background has not been explicitly discussed, yet.
We thus want to determine the parameters $\{\theta_{i}\}$ only by OA
as is established in the case of 2D Ising model~\cite{Onsager}. 
However, the representation theory of OA has not been fully developed. 
For instance, all the finite-dimensional representations are not 
classified, yet.  
On the other hand, it is known that 
the OA is isomorphic to a subalgebra of 
the $\mathfrak{sl}_{2}$-loop algebra, 
$L(\mathfrak{sl}_{2})$~\cite{Date-Roan}. 
It is thus natural to ask whether 
the polynomial $P_{\mathrm{CP}}(\zeta)$
can be understood in terms of the representation theory of 
$L(\mathfrak{sl}_{2})$~\cite{Drinfeld,Chari-Pressley}. 
Our main purpose here is to discuss 
the polynomial $P_{\mathrm{CP}}(\zeta)$
from an algebraic point of view. The result might be useful  
to develop the representation theory 
of OA in order to determine the parameters $\{\theta_{i}\}$.  

Bazhanov and Stroganov introduced 
an integrable $N$-state spin chain,  
which connects the chiral Potts model 
to the six-vertex model~\cite{Bazhanov-Stroganov}.
At the superintegrable point, the model 
is called nilpotent Bazhanov-Stroganov (NBS) model 
and is equivalent to an XXZ-type spin chain at $q^{N}=1$.   
It has therefore Bethe eigenstates. 
The NBS transfer matrix $\tau_{\mathrm{NBS}}(z)$ 
commutes with the transfer matrix of the SCP model 
as well as with the von Gehlen-Rittenberg's Hamiltonian 
$H_{\mathrm{vGR}}$~\cite{Bazhanov-Stroganov}.     
However, not all the Bethe states of $\tau_{\mathrm{NBS}}(z)$ 
are eigenstates of $H_{\mathrm{vGR}}$: 
as shown in~\cite{Baxter,Tarasov},  from a given Bethe 
eigenstate of $\tau_{\mathrm{NBS}}(z)$, the Hamiltonian 
 $H_{\mathrm{vGR}}$ generates such a subspace 
that has the Ising-like spectra \eqref{eq:energy-form_SCP}, 
while it is nontrivial to construct a complete set of 
eigenvectors of $H_{\mathrm{vGR}}$ in the subspace. Furthermore, 
the subspace gives a degenerate eigenspace of 
$\tau_{\mathrm{NBS}}(z)$ with respect to the OA symmetry.    
Here we note that the commutativity 
$[H_{\mathrm{vGR}}, \tau_{\mathrm{NBS}}(z)]=0$ 
leads to the OA symmetry of the NBS model.  

We now discuss the Ising-like spectra of the SCP model 
from the viewpoint of the $L(\mathfrak{sl}_{2})$ symmetry.  
Recently it is found that the XXZ spin chain at $q^{N}=1$ has 
large degeneracies in the energy spectra and 
the degeneracies 
are described by the $L(\mathfrak{sl}_{2})$ 
symmetry~\cite{Deguchi-Fabricius-McCoy}. 
In this letter, we show that the NBS model   
has the $L(\mathfrak{sl}_{2})$ symmetry in a certain sector.  
It indeed gives a ``higher'' symmetry than the OA symmetry.  
Applying the approach of \cite{Deguchi_cond-mat} 
to a Bethe state of $\tau_{\mathrm{NBS}}(z)$ in the sector, 
we obtain the Drinfeld polynomial 
if the zeros are distinct \cite{Chari-Pressley2}. 
We then find that the Drinfeld polynomial 
is equivalent to the chiral Potts polynomial $P_{\mathrm{CP}}(\zeta)$
of the Ising-like spectra associated with the Bethe state. 
Therefore, the representation space of OA for the polynomial 
$P_{\mathrm{CP}}(\zeta)$ has the same dimensions as 
the $L(\mathfrak{sl}_{2})$-degenerate eigenspace of 
the NBS model $\tau_{\mathrm{NBS}}(z)$. 

The letter consists of the following: 
in \S 2 we introduce the NBS model; in \S 3 we derive the 
 $L(\mathfrak{sl}_{2})$ symmetry of the NBS model in a sector; 
in \S 4 we show that the Drinfeld polynomial is equivalent to  
the chiral Potts polynomial $P_{\mathrm{CP}}(\zeta)$ in the sector, 
and finally we give a conjecture.

\mathversion{bold}
\section{Nilpotent Bazhanov-Stroganov model}
\mathversion{normal}

First we introduce an XXZ-type spin chain defined for generic $q$.
Later the model is identified with the NBS model
when $q^{N}=1$. 
The $L$-operators $\mathcal{L}_{i}(z)\in
\End(\mathbb{C}^{2}\otimes(\mathbb{C}^{N})^{\otimes L}),
(i=1,\ldots,L)$ for the model are given by
\begin{align}
\label{eq:L-op_XXZ}
 &\mathcal{L}_{i}(z)\Define
  \begin{pmatrix}
   q^{-\frac{1}{2}}
   (z\Hat{k}_{i}^{\frac{1}{2}}-z^{-1}\Hat{k}_{i}^{-\frac{1}{2}})
   & (q-q^{-1})\Hat{f}_{i} \\[-1mm]
   (q-q^{-1})\Hat{e}_{i}
   &q^{\frac{1}{2}}
   (z\Hat{k}_{i}^{-\frac{1}{2}}-z^{-1}\Hat{k}_{i}^{\frac{1}{2}}) \\
  \end{pmatrix}
  .
\end{align}
Here $\{\Hat{k}_{i}, \Hat{e}_{i},\Hat{f}_{i}\}$ is
the $N$-dimensional irreducible
representation of the quantum group
$U_{q}(\mathfrak{sl}_{2})$
acting on the $i$th component of 
the tensor product $(\mathbb{C}^{N})^{\otimes L}$.
We construct the monodromy matrix 
$\mathcal{T}(z)$ and the transfer matrix $\tau(z)$ as 
\begin{align}
 \mathcal{T}(z)\Define
 \overset{\curvearrowleft}{\prod_{i=1}^{L}}
 \mathcal{L}_{i}(z)
 \definE
 \begin{pmatrix}
  A(z) & B(z) \\[-1mm]
  C(z) & D(z)
 \end{pmatrix}
 ,\quad
 \tau(z)\Define 
 \mathrm{tr}_{\mathbb{C}^{2}}(\mathcal{T}(z))
 =A(z)+D(z), \nn
\end{align}
where $A(z),B(z),C(z),D(z)
\in\End((\mathbb{C}^{N})^{\otimes L})$.
It is well-known that the transfer matrix $\tau(z)$ forms 
a commutative family, $\tau(z)\tau(w)=\tau(w)\tau(z)$.
Since the $\mathcal{T}(z)$ is intertwined by the $R$-matrix
of the six-vertex model, the operators $A(z),B(z),C(z)$ 
and $D(z)$ satisfy the same relations as those
in the case of the spin-$1/2$ XXZ spin 
chain~\cite{Korepin-Bogoliubov}.
Then it is straightforward to apply the algebraic Bethe ansatz
to the XXZ-type spin chain~\eqref{eq:L-op_XXZ}.
Let $|0\rangle\Define v_{0}\otimes\dots\otimes v_{0}
\in(\mathbb{C}^{N})^{\otimes L}$ be the reference state.
One sees
\begin{align}
&A(z)|0\rangle
 =q^{-\frac{L}{2}}(zq^{\frac{N\!-\!1}{2}}\!\!
  -\!z^{-1}q^{-\frac{N\!-\!1}{2}})^{L}|0\rangle,\quad
 D(z)|0\rangle
 =q^{\frac{L}{2}}(zq^{-\frac{N\!-\!1}{2}}\!\!
 -\!z^{-1}q^{\frac{N\!-\!1}{2}})^{L}|0\rangle,\nn\\
&C(z)|0\rangle=0.\nn
\end{align}
If a set of variables
$\{z_{i}|i=1,\ldots,R\}$ satisfies the Bethe equations
\begin{align}
 \label{eq:Bethe-eq_BS}
 \left(\frac{z_{i}^{2}q^{N-1}-1}{z_{i}^{2}-q^{N-1}}\right)^{L}
 =q^{L}\prod_{j=1 \atop j(\neq i)}^{R}
  \frac{z_{i}^{2}q^{2}-z_{j}^{2}}{z_{i}^{2}-z_{j}^{2}q^{2}},
\end{align}
the Bethe state
$|R;\{z_{i}\}\rangle\Define\prod_{i=1}^{R}B(z_{i})|0\rangle$
gives an eigenstate of the transfer matrix $\tau(z)$.

Secondly 
we see that, at $q^{N}=1$, the XXZ-type spin chain 
\eqref{eq:L-op_XXZ} reduces to the NBS model,
i.e., $\tau(z)|_{q^{N}=1}=\tau_{\mathrm{NBS}}(z)$.
The $L$-operators \eqref{eq:L-op_XXZ} are rewritten as 
\begin{align}
\label{eq:L-op_BS}
&\mathcal{L}_{i}(z)=
  \begin{pmatrix}
   -zq^{-1}Z_{i}^{-1}+z^{-1}Z_{i} 
   & -(Z_{i}^{-1}-Z_{i})X_{i} \\[-1mm]
   X_{i}^{-1}(Z_{i}^{-1}-Z_{i})
   &z^{-1}Z_{i}^{-1}-zqZ_{i} \\
  \end{pmatrix}
  ,
\end{align} 
through the following
nilpotent representation of $U_{q}(\mathfrak{sl}_{2})$
at $q^{N}=1$:
\begin{align}
\label{eq:nilpotent-rep}
 \Hat{k}^{\frac{1}{2}}_{i}=-q^{-\frac{1}{2}}Z_{i}^{-1},\quad
 \Hat{e}_{i}=
 X_{i}^{-1}\frac{Z_{i}^{-1}-Z_{i}}{q-q^{-1}},\quad
 \Hat{f}_{i}=
 \frac{Z_{i}-Z_{i}^{-1}}{q-q^{-1}}X_{i}.
\end{align}
Recall that the $Z_{i}$ and $X_{i}$ are the operators 
defined at $q^{N}=1$ in \eqref{eq:ZX-op}.
The $L$-operators \eqref{eq:L-op_BS} 
are equivalent to the original ones%
~\cite{Bazhanov-Stroganov} at superintegrable point: 
we take the principal gradation~\cite{Belavin-Odesskii-Usmanov}
and make an appropriate similarity transformation, then we obtain 
the expression \eqref{eq:L-op_BS}.
Since the representation \eqref{eq:nilpotent-rep} 
is called nilpotent in contrast to the cyclic representation of 
$U_{q}(\mathfrak{sl}_{2})$ at $q^{N}=1$, 
we have referred to the model as the nilpotent BS model. 
One also notices that, through the change of variables 
$z_{i}\mapsto -q^{3}u_{i}$,
the Bethe equations~\eqref{eq:Bethe-eq_BS} are identified with
the coupled nonlinear equations \eqref{eq:CP-Bethe-eq}
with $q^{N}=1$ and some specific $p_{a}$ and $p_{b}$.
In fact, at $q^{N}=1$, the Bethe state $|R;\{z_{i}\}\rangle$
is shown to belong to the subspace with the spectra
\eqref{eq:energy-form_SCP} characterized by
$P_{\mathrm{CP}}(\zeta)$~\eqref{eq:CP-poly}~\cite{Tarasov}.

\mathversion{bold}
\section{Loop algebra $L(\mathfrak{sl}_{2})$ symmetry}
\mathversion{normal}

We show that the NBS model has a loop algebra 
$L(\mathfrak{sl}_{2})$ symmetry. 
We first consider the operators $A(z), B(z), C(z)$ and $D(z)$
with generic $q$ and then take the limit $q^{N}\to 1$ to discuss
the NBS model.
Introduce 
\begin{align}
&A\Define\lim_{z\to\infty}\frac{A(z)}{z^{L}q^{-\frac{L}{2}}}
 =\lim_{z\to\infty}\frac{D(z)}{z^{-L}q^{\frac{L}{2}}}
 =\underbrace{\Hat{k}^{\frac{1}{2}}\otimes\cdots
 \otimes \Hat{k}^{\frac{1}{2}}}_{L}, \nn\\
&B_{\pm}\Define\lim_{z^{\pm 1}\to\infty}
 \frac{B(z)}{n_{\pm}(z)}
 =q^{-\frac{1}{2}(L+1)}
 \sum_{i=1}^{L}q^{i}
 \underbrace{\Hat{k}^{\mp\frac{1}{2}}\otimes\cdots
 \otimes\Hat{k}^{\mp\frac{1}{2}}}_{i-1}
 \otimes\Hat{f}\otimes
 \underbrace{\Hat{k}^{\pm\frac{1}{2}}\otimes\cdots
 \otimes\Hat{k}^{\pm\frac{1}{2}}}_{L-i},\nn\\
&C_{\pm}\Define\lim_{z^{\pm 1}\to\infty}
 \frac{C(z)}{n_{\pm}(z)}
 =q^{\frac{1}{2}(L+1)}
 \sum_{i=1}^{L}q^{-i}
 \underbrace{\Hat{k}^{\pm\frac{1}{2}}\otimes\cdots
 \otimes\Hat{k}^{\pm\frac{1}{2}}}_{i-1}
 \otimes\Hat{e}\otimes
 \underbrace{\Hat{k}^{\mp\frac{1}{2}}\otimes\cdots
 \otimes\Hat{k}^{\mp\frac{1}{2}}}_{L-i},\nn
\end{align}
with normalization factors 
$n_{\pm}(z)=(\pm z^{\pm1})^{L-1}(q-q^{-1})$.
Through the relations among $A(z), B(z), C(z)$ and $D(z)$,
we find that the operators
\[
 k_{0}^{-1}=k_{1}=A^{2},\quad
 e_{0}=B_{+},\quad
 e_{1}=C_{+},\quad
 f_{0}=C_{-},\quad
 f_{1}=B_{-},
\]
give a finite-dimensional representation of the
quantum affine algebra $U_{q}^{\prime}(\Hat{\mathfrak{sl}}_{2})$.
After a simple calculation, we obtain
\begin{align}
 (B_{\pm})^{m}
&=\!\!\!
 \sum_{0\leq\lambda_{i}\leq m \atop 
 \lambda_{1}+\cdots+\lambda_{L}=m}\!\!\!
 q^{\sum_{j}\left(j-\frac{1}{2}(L+1)\right)\lambda_{j}}
 \frac{[m]!}{[\lambda_{1}]!\cdots [\lambda_{L}]!}\;
 \bigotimes_{i=1}^{L}\;
 \Hat{f}^{\lambda_{i}}
 \Hat{k}^{\pm\frac{1}{2}
 \left(\sum_{j(<i)}\!-\!\sum_{j(>i)}\right)\lambda_{j}},
 \nn\\
 (C_{\pm})^{m}
&=\!\!\!
 \sum_{0\leq\lambda_{i}\leq m \atop 
 \lambda_{1}+\cdots+\lambda_{L}=m}\!\!\!
 q^{-\sum_{j}\left(j-\frac{1}{2}(L+1)\right)\lambda_{j}}
 \frac{[m]!}{[\lambda_{1}]!\cdots [\lambda_{L}]!}\;
 \bigotimes_{i=1}^{L}\;
 \Hat{k}^{\mp\frac{1}{2}
 \left(\sum_{j(<i)}\!-\!\sum_{j(>i)}\right)\lambda_{j}}
 \Hat{e}^{\lambda_{i}},\nn
\end{align}
where $[m]\Define\frac{q^{m}-q^{-m}}{q-q^{-1}}$
and $[m]!\Define\prod_{i=1}^{m}[i]$.

We now consider the limit $q^{N}\to 1$.
One easily finds from the expression above that, 
if we set $q^{N}=1$, then
$(B_{\pm})^{N}=(C_{\pm})^{N}=0$.
Define
\begin{align}
\label{eq:HBC^(n)}
&H^{(n)}
 \Define\frac{1}{n}
 \sum_{i=1}^{L}\underbrace{\id\otimes\cdots\otimes\id}_{i-1}
 \otimes\,\Hat{h}
 \otimes\underbrace{\id\otimes\cdots\otimes\id}_{L-i},
 \nn\\
&B_{\pm}^{(n)}\Define\lim_{q^{N}\to 1}
 \frac{(B_{\pm})^{n}}{[n]!},\qquad
 C_{\pm}^{(n)}\Define\lim_{q^{N}\to 1}
 \frac{(C_{\pm})^{n}}{[n]!},
\end{align}
where $\Hat{h}=\mathrm{diag}\{N-1,N-3,\ldots,-N+1\}
\in\End(\mathbb{C}^{N})$.
By considering the relations among 
$A(z), B(z), C(z)$ and $D(z)$ in the limit $q^{N}\to 1$, we have
\begin{align}
 &[\tau_{\mathrm{NBS}}(z),B_{\pm}^{(N)}]
  =-z^{\pm 1}B_{\pm}^{(N-1)}B(z)
   (q^{-\frac{L}{2}}A^{\pm 1}
   -q^{\frac{L}{2}}A^{\mp 1}), \nn\\
 &[\tau_{\mathrm{NBS}}(z),C_{\pm}^{(N)}]
  =z^{\pm 1}C_{\pm}^{(N-1)}C(z)
   (q^{-\frac{L}{2}}A^{\pm 1}
   -q^{\frac{L}{2}}A^{\mp 1}). \nn
\end{align}
Recall that $\tau(z)|_{q^{N}=1}=\tau_{\mathrm{NBS}}(z)$.
Then we have 
$[\tau_{\mathrm{NBS}}(z),B_{+}^{(N)}]
=[\tau_{\mathrm{NBS}}(z),C_{+}^{(N)}]=0$ in the sector
$A^{2}=q^{L}$ and
$[\tau_{\mathrm{NBS}}(z),B_{-}^{(N)}]
=[\tau_{\mathrm{NBS}}(z),C_{-}^{(N)}]=0$ in the sector
$A^{2}=q^{-L}$.

From the relations among $\{A,B_{\pm},C_{\pm}\}$
and the general theory of quantum affine algebras 
at roots of unity~\cite{Lusztig}, we show
\begin{align}
 &[B_{+}^{(N)},B_{-}^{(N)}]=
  [C_{+}^{(N)},C_{-}^{(N)}]=0,\nn\\
 &[H^{(N)},B_{\pm}^{(N)}]=-2B_{\pm}^{(N)},\qquad
  [H^{(N)},C_{\pm}^{(N)}]=2C_{\pm}^{(N)},\nn\\
 &[B_{\pm}^{(N)},[B_{\pm}^{(N)},
  [B_{\pm}^{(N)},C_{\pm}^{(N)}]]]=0,\qquad
  [C_{\pm}^{(N)},[C_{\pm}^{(N)},
  [C_{\pm}^{(N)},B_{\pm}^{(N)}]]]=0. \nn
\end{align}
Furthermore, in the sector with $A^{2}=1$ 
in $(\mathbb{C}^{N})^{\otimes L}$, we have
\[
  [B_{\pm}^{(N)},C_{\mp}^{(N)}]=\pm H^{(N)}.
\]
Let $\{H_{i},E_{i},F_{i}|i=0,1\}$ express
the Chevalley generators of 
the $\mathfrak{sl}_{2}$-loop algebra
$L(\mathfrak{sl}_{2})$.
The map $\Hat{\;}:L(\mathfrak{sl}_{2})\to
\End((\mathbb{C}^{N})^{\otimes L})$ defined by
\[
 \Hat{H}_{0}=-\Hat{H}_{1}\Define -H^{(N)},\quad
 \Hat{E}_{0}\Define B_{+}^{(N)},\quad
 \Hat{E}_{1}\Define C_{+}^{(N)},\quad
 \Hat{F}_{0}\Define C_{-}^{(N)},\quad
 \Hat{F}_{1}\Define B_{-}^{(N)},
\]
gives a finite-dimensional representation of the loop algebra 
$L(\mathfrak{sl}_{2})$ in the sector with $A^{2}=1$.
Thus we have the following: 
\begin{prop}[cf.~\cite{Deguchi-Fabricius-McCoy}]
If $q^{N}=1$ and $L$ is a multiple of $N$, 
the transfer matrix $\tau_{\mathrm{NBS}}(z)$ has 
the $L(\mathfrak{sl}_{2})$ symmetry
in the sector with $A^{2}=1$:
\[
  [\tau_{\mathrm{NBS}}(z),\Hat{H}_{0,1}]
  =[\tau_{\mathrm{NBS}}(z),\Hat{E}_{0,1}]
  =[\tau_{\mathrm{NBS}}(z),\Hat{F}_{0,1}]=0.
\]
\end{prop}

\section{$L(\mathfrak{sl}_{2})$-degenerate eigenspaces 
and the SCP spectra}

We now discuss the dimensions of $L(\mathfrak{sl}_{2})$-degenerate 
eigenspaces of the NBS model. 
Let us express the highest weight conditions of the 
Drinfeld realization of $L(\mathfrak{sl}_{2})$ 
in terms of the Chevalley generators \cite{Drinfeld,Chari-Pressley}.  
We call a vector $\Omega$ a highest weight vector 
if it is annihilated by $E_{1}$ and $F_{0}$,
$E_{1}\Omega=F_{0}\Omega=0$,
and is diagonalized by $H_{1}(=-H_{0})$, $(E_{1})^{k}(E_{0})^{k}$
and $(F_{0})^{k}(F_{1})^{k}$ for $k\in\mathbb{Z}_{\geq 0}$.
If a representation is generated 
by a highest weight vector, we call it highest weight.  
For a finite-dimensional 
representation generated by highest weight vector $\Omega$,  
we define the following polynomial:
\[
  P_{\mathrm{D}}(\xi)=\sum_{k=0}^{n}\lambda_{k}(-\xi)^{k} , 
\]
where $n$ and $\lambda_{k}$ denote the eigenvalues of
$H_{1}$ and $(E_{1})^{k}(E_{0})^{k}/(k!)^{2}$ on $\Omega$, 
respectively. (See eq.~(4.2) of \cite{Fabricius-McCoy}.) 
If the zeros of  $P_{\mathrm{D}}(\xi)$ are distinct, the 
representation is irreducible \cite{Chari-Pressley2}. 
And the polynomial $P_{\mathrm{D}}(\xi)$  is called 
the Drinfeld polynomial.  

Consider the Bethe states of the NBS model 
in the sector with $A^{2}=q^{-L-2R}= 1$. 
Here we recall that both $L$ and $R$ are multiples of $N$. 
Let $\{z_{i}\}$ be a set of regular solutions 
of the Bethe equations~\eqref{eq:Bethe-eq_BS} at $q^{N}=1$. 
Here, if solutions of ~\eqref{eq:Bethe-eq_BS}, $\{z_{i}\}$, 
are finite, distinct and nonzero, we call them regular.  
In the similar way to~\cite{Deguchi_cond-mat}, it is shown that
the Bethe state $|R;\{z_{i}\} \rangle$ is a highest weight vector. 
After some calculation, we obtain
\begin{align}
&\Hat{H}_{1}|R;\{z_{i}\}\rangle
 =\frac{L(N-1)-2R}{N}|R;\{z_{i}\}\rangle, \nn\\
&\frac{(\Hat{E}_{1})^{k}}{k!}
 \frac{(\Hat{E}_{0})^{k}}{k!}|R;\{z_{i}\}\rangle
 =(-)^{k}\chi_{kN}|R;\{z_{i}\}\rangle,\qquad
 \left(0\leq k\leq\frac{L(N-1)-2R}{N}\right), \nn
\end{align}
where $\chi_{k}$ is defined 
by the following series expansion: 
\begin{align}
 &\frac{\prod_{i=1}^{N-1}\phi(xq^{2i-N})}
  {F(xq)F(xq^{-1})}
  =\frac{(1-x^{N})^{L}}
  {(1-x)^{L}F(xq)F(xq^{-1})}
  =\sum_{k=0}^{\infty}\chi_{k}x^{k} . 
\end{align}
The functions $\phi(x)$ and $F(\xi)$ are given by  
$\phi(x)\Define (1-x)^{L}$ 
and $F(\xi)\Define\prod_{i=1}^{R} (1-\xi z_{i})$, respectively. 
The result is summarized as follows: 

\begin{prop}
If $L$ and $R$ are  multiples of $N$, the Bethe state 
of the NBS model, $|R;\{z_{i}\}\rangle$, is highest weight. 
The polynomial $P_{\mathrm{D}}(\zeta)$ 
of the finite-dimensional representation 
generated by $|R;\{z_{i}\}\rangle$ is given by 
\begin{align}
\label{eq:Drinfeld-poly}
&P_{\mathrm{D}}(\xi^{N})
 =\!\!\sum_{k=0}^{\frac{L(N-1)-2R}{N}}\!\!\chi_{kN}\xi^{kN}
 =N\sum_{j=0}^{N-1}
  \frac{(1-\xi^{N})^{L}}
  {(1-\xi q^{2j})^{L}F(\xi q^{2j-1})F(\xi q^{2j+1})}. 
\end{align}
\end{prop}

The polynomial $P_{\mathrm{D}}(\zeta)$ gives 
the Drinfeld polynomial if the zeros are distinct.
In general, if the Drinfeld polynomial of a finite-dimensional 
irreducible representation is factorized as 
$P_{\mathrm{D}}(\zeta)
=\prod_{i=1}^{n}(1-\zeta_{i}^{-1}\zeta)^{l_{i}}$
with distinct zeros $\zeta_{1},\ldots,\zeta_{n}\in\mathbb{C}$, 
the dimensions of the representation 
are given by $\prod_{i=1}^{n}(l_{i}+1)$. 
Thus, in the case $l_{i}=1$ for all $i$, 
the dimensionality of the $L(\mathfrak{sl}_{2})$-degenerate 
eigenspace of $\tau_{\mathrm{NBS}}(z)$ 
is $2^{\mathrm{deg} P_{\mathrm{D}}(\zeta)}$. 

Comparing expression \eqref{eq:CP-poly} 
with \eqref{eq:Drinfeld-poly} we have the following: 
\begin{prop} If the zeros of the chiral Potts polynomial 
$P_{\mathrm{CP}}(\zeta)$ \eqref{eq:CP-poly} are distinct, 
the $P_{\mathrm{CP}}(\zeta)$
with $p_{a}+p_{b}=0$ is equivalent
to the Drinfeld polynomial $P_{\mathrm{D}}(\zeta)$ 
in \eqref{eq:Drinfeld-poly}.
\label{coinc}
\end{prop} 

It is suggested from proposition \ref{coinc} 
that the $2^{\mathrm{deg} P_{\mathrm{CP}}(\zeta)}$-dimensional 
representation space of OA characterized by the polynomial 
$P_{\mathrm{CP}}(\zeta)$ 
corresponds to the degenerate eigenspace of $L(\mathfrak{sl}_{2})$ 
of the Drinfeld polynomial $P_{\mathrm{D}}(\zeta)$. 
Here we remark that we have verified this conjecture 
in the case of $L=N=3$. 

The representation space of OA of the chiral Potts polynomial 
$P_{\mathrm{CP}}(\zeta)$ and 
the $L(\mathfrak{sl}_{2})$-degenerate eigenspace of 
the Drinfeld polynomial $P_{\mathrm{D}}(\zeta)$ 
have the same dimensions. 
Furthermore, they have the same Bethe state. 
As is shown in~\cite{Baxter,Tarasov}, the subspace characterized 
by $P_{\mathrm{CP}}(\zeta)$ is generated
by repeated application of the SCP transfer matrix to 
the Bethe state. 
The conjecture is derived if we assume that Bethe states 
of $\tau_{\mathrm{NBS}}(z)$ are complete and 
also that the Bethe states are nondegenerate with respect to 
eigenvalues of $\tau_{\mathrm{NBS}}(z)$.  

We have clarified 
one of the fundamental algebraic aspects of the polynomial 
$P_{\mathrm{CP}}(\zeta)$ 
characterizing the Ising-like spectrum of SCP model.
One notices that the $L(\mathfrak{sl}_{2})$ symmetry of 
the NBS model readily provides an OA symmetry of the model.
We thus speculate that, in a certain sector, the OA symmetry 
should be the origin of the OA structure of 
both SCP model and von Gehlen-Rittenberg's model $H_{\mathrm{vGR}}$. 
However, it is our future problem to discuss the representation of 
$A_{0}$ and $A_{1}$ in terms of 
the $L(\mathfrak{sl}_{2})$ representation.

\section*{Acknowledgments}

The authors would like to thank Prof.~A.~Kuniba 
and Prof.~N.~Hatano for helpful comments.
One of the authors (AN) appreciates the Research Fellowships of the
Japan Society for the Promotion of Science for Young Scientists. 
The present study  is partially supported by 
Grant-in-Aid for Scientific Research (C) No. 17540351.


\end{document}